\documentclass[conference,letterpaper]{IEEEtran}
\usepackage[T1]{fontenc}
\usepackage[utf8]{inputenc}
\usepackage{cite}
\usepackage{amsmath,amssymb}
\usepackage{graphicx}
\usepackage{booktabs,array,tabularx}
\usepackage[caption=false,font=footnotesize]{subfig}
\usepackage{url}
\usepackage{flushend}
\usepackage[hidelinks]{hyperref}
\graphicspath{{figures/}}
\begin{document}
\title{Performance Benchmarking of Federated Learning Frameworks for Medical Image Classification}
\author{
\IEEEauthorblockN{Vithor Pileco}
\IEEEauthorblockA{Institute of Informatics\\UFRGS\\Porto Alegre, Brazil\\vithor.pileco@inf.ufrgs.br}
\and
\IEEEauthorblockN{Luan Mantegazine}
\IEEEauthorblockA{Institute of Informatics\\UFRGS\\Porto Alegre, Brazil\\luan.mantegazine@inf.ufrgs.br}
\and
\IEEEauthorblockN{Claudio Geyer}
\IEEEauthorblockA{Institute of Informatics\\UFRGS\\Porto Alegre, Brazil\\geyer@inf.ufrgs.br}
}
\maketitle

\begin{abstract}
Selecting a Federated Learning (FL) framework requires understanding the resource demands of its training and evaluation workflow. This work compares Flower, NVIDIA FLARE, and FedBioMed using two-dimensional MedNIST classification with ResNet-10 in a containerized, single-host environment. Following a dated review of repository activity and documented medical applications, we evaluated nine configurations per framework, with three random seeds, for a total of 81 executions. The configurations varied federation size, communication rounds, and batch size. We monitored execution time, CPU and RAM consumption, host-level GPU utilization, server network traffic rates, and model accuracy at the end of each execution. The implementations shared nominal learning settings while retaining differences in data splitting and federated evaluation. Configuration-blocked non-parametric analysis identified differences in execution time, client and server memory, GPU utilization, and network traffic rates. NVIDIA FLARE exhibited the lowest server-memory consumption and the highest host-level GPU utilization, Flower exhibited the highest average server network traffic rate, and FedBioMed required more client memory and longer execution time. No statistically significant difference in final accuracy was detected within the evaluated short training schedules; this does not establish convergence or equivalent predictive performance. The results characterize the configured workflows under shared hardware resources. They provide a concrete case study of system behavior, with conclusions bounded by workflow asymmetries, local networking, and the lightweight MedNIST workload.
\end{abstract}


\section{Introduction}
\label{sec:introduction}
Artificial Intelligence (AI) has become increasingly relevant to medical image analysis, supporting applications such as automated screening and disease detection \cite{Topol2019}, \cite{McKinney2020}. Developing models for clinical environments nevertheless requires attention to differences between institutions, patient populations, and acquisition settings. Models developed using data from one institution may exhibit reduced generalization when evaluated in other settings \cite{Avendano2024}, \cite{Yala2021}. Federated Learning (FL) enables participating institutions to train a shared model while retaining their datasets locally \cite{McMahan2017}. In a typical FL workflow, a server coordinates clients, distributes the global model, and aggregates locally trained updates over successive communication rounds.

Selecting an FL software framework is also an engineering decision. A recent survey identified 29 open-source FL frameworks \cite{Boekhoff2025}, with different architectures, supported algorithms, and deployment interfaces. These choices affect the implementation of training and evaluation workflows and can influence execution time, memory use, and communication behavior. Measuring these quantities is useful when research teams must choose software that fits the resources available for an experiment.

Medical imaging provides a relevant application context because data handling, integration with image-processing pipelines, and available computational resources influence the feasibility of collaborative training. The present study addresses one bounded part of this problem: the behavior of FL software when executing a lightweight two-dimensional image-classification task. We use MedNIST and ResNet-10 to examine configured workflows within a common hardware environment. The empirical contribution is the joint characterization of the implemented workflows and their resource behavior; its applicability to volumetric CT or MRI remains to be evaluated.

Existing studies provide software-oriented comparisons and experimental evaluations of FL frameworks. Our contribution is a complementary systems evaluation that combines continuous resource monitoring with controlled variation of federation size, communication rounds, and batch size. The comparison includes Flower, NVIDIA FLARE, and FedBioMed, with separate measurements of client and server resources where supported by the instrumentation. Section II positions this combination relative to previous benchmarks.

Framework selection began with 30 candidates identified from previous studies and complementary sources. Applying a repository-activity criterion at a specified date identified seven candidates, from which Flower, NVIDIA FLARE, and FedBioMed were selected on the basis of documented medical applications. We implemented MedNIST classification \cite{Cardoso2022} using the same model architecture, optimizer, nominal training settings, and FedAvg aggregation in all three frameworks. The implementations retained framework-specific differences, particularly in train/test splitting and the execution of federated evaluation. The benchmark therefore compares configured native workflows; its results do not isolate framework internals from those workflow differences.

The study comprises nine configurations per framework and three executions per configuration, using common seed values, for a total of 81 executions. All client and server containers share one physical host. We measure execution time, CPU and RAM consumption, host-level GPU utilization, server network traffic rates, and accuracy after each short training schedule. Increasing the number of clients in this environment changes resource contention as well as orchestration and communication demands. Accuracy is reported as an outcome of the evaluated budget, without claiming convergence or equivalent predictive performance across frameworks.

The main contributions are:
\begin{itemize}
\item A dated characterization of the FL framework ecosystem using an explicit repository-activity criterion and documented medical applications to select the evaluated frameworks.
\item A containerized MedNIST/ResNet-10 case study covering 81 executions of Flower, NVIDIA FLARE, and FedBioMed across variations in federation size, communication rounds, and batch size.
\item Continuous measurement of CPU, RAM, GPU utilization, and network traffic rates alongside execution time and model accuracy, with client/server separation where supported by the instrumentation.
\item A configuration-blocked, non-parametric comparison of the observed outcomes, together with descriptive analysis of parameter-dependent behavior under shared hardware resources.
\item An account of workflow differences and measurement limitations that bounds the interpretation of the results and informs their use when planning subsequent FL experiments.
\end{itemize}

The remainder of this paper is organized as follows. Section II discusses related work. Section III describes the experimental methodology. Section IV presents and discusses the results. Section V concludes the paper and outlines its limitations and directions for further evaluation.

\section{Related Work}
\label{sec:related_work}

FL frameworks have been compared in terms of functionality, usability, learning outcomes, and systems performance. Boekhoff et al.~\cite{Boekhoff2025} surveyed the open-source ecosystem, while Chavero-Diez et al.~\cite{Chavero2026} examined software quality and interoperability for biomedical research. These studies inform framework selection and motivate explicit reporting of the scope and configurations of empirical comparisons.

Several studies provide experimental benchmarking. Liu et al.~\cite{Liu2023} introduced UniFed to evaluate multiple FL frameworks across model- and system-level dimensions. Soudan et al.~\cite{Soudan2025} compared Flower, FedML, Substra, and OpenFL while varying client count and measuring accuracy, training time, CPU utilization, and RAM consumption. Batsios and Angelis~\cite{Batsios2026} compared six frameworks under controlled FedAvg settings in a virtualized testbed. Tapia et al.~\cite{Tapia2026} considered FL frameworks in discrete-event manufacturing systems. These works establish that systems benchmarking is an existing research direction; the present study contributes an additional, explicitly bounded empirical comparison.

Within medical imaging, Fonio~\cite{Fonio2024} examined four frameworks and investigated strong scaling by increasing client count while keeping the total dataset fixed, with an emphasis on execution time and scalability. Gupta et al.~\cite{Gupta2025} evaluated NVIDIA FLARE, Flower, and Owkin Substra on PathMNIST. Their report considers learning outcomes, communication, computational resources, and deployment characteristics for a five-client setup. The present study complements that comparison by including FedBioMed and quantifying CPU, RAM, host GPU utilization, and network rates under variations of client count, round count, and batch size. It also applies configuration-blocked inference to the measured outcomes. These differences concern the framework set, instrumentation, and experimental design; using another medical dataset alone does not establish novelty. Results from the two studies are not directly interchangeable because their models, workloads, settings, and measurements differ.

FedBed, proposed by Symeonides et al.~\cite{Symeonides2023}, provides configurable virtual testbeds for FL workflows, including control over computational resources, networking, and data distributions. It addresses infrastructure-aware experimentation across edge environments. Our study addresses a narrower question: how three configured FL implementations behave on a fixed host under a common MedNIST task. It does not introduce a competing testbed abstraction or evaluate the infrastructure heterogeneity supported by FedBed.

Table~\ref{tab:related_work} summarizes this relationship to selected benchmarks. The contribution is the combination of a specific framework set, continuous resource measurements, parameter variation, and a blocked comparison of the observed executions. Medical imaging supplies the use case and motivates software selection, while the empirical claims remain limited to lightweight two-dimensional classification.

\begin{table*}[t]
\centering
\small
\caption{Positioning relative to selected experimental FL benchmarks. The comparison describes study scope rather than a ranking of benchmark quality.}
\label{tab:related_work}
\renewcommand{\arraystretch}{1.10}
\begin{tabularx}{\textwidth}{@{}>{\raggedright\arraybackslash}p{0.14\textwidth}>{\raggedright\arraybackslash}p{0.36\textwidth}>{\raggedright\arraybackslash}X@{}}
\toprule
Study & Main scope & Relationship to this study \\
\midrule
Liu et al.~\cite{Liu2023} & Unified evaluation of multiple FL frameworks. & Establishes a broader framework-benchmarking context. \\
Soudan et al.~\cite{Soudan2025} & Framework comparison with client-count variation and runtime/resource measurements. & Shares the systems-comparison objective; the present case adds its selected medical workload and framework set. \\
Fonio~\cite{Fonio2024} & Medical-image classification and strong scaling with fixed total data. & Shares the application context and client-count dimension. \\
Gupta et al.~\cite{Gupta2025} & FLARE, Flower, and Substra on PathMNIST, including resource and deployment considerations. & Closest application-oriented comparison; this study includes FedBioMed, continuous monitoring, and three parameter dimensions. \\
Symeonides et al.~\cite{Symeonides2023} & FedBed: configurable virtualized FL testbeds with resource, network, and data-distribution controls. & Addresses configurable infrastructure; this study characterizes fixed-host implementations. \\
This work & Flower, FLARE, and FedBioMed on MedNIST/ResNet-10, with 81 executions. & Quantitative resource profiles and configuration-blocked comparisons, bounded by shared hardware and workflow asymmetries. \\
\bottomrule
\end{tabularx}
\end{table*}

\section{Experimental Setup}
\label{sec:experimental_setup}

We compared Flower, NVIDIA FLARE, and FedBioMed using a common dataset, model architecture, optimizer, and nominal training schedule. The benchmark measures the complete configurations implemented in each framework, including their framework-specific training and evaluation behavior. Each client and server ran in a separate Docker container. Differences in data splitting, evaluation messages, and training APIs are documented below because they limit the equivalence of the measured workloads.

\subsection{Framework Selection}

Starting from 30 frameworks identified in previous studies and complementary sources, we used repository activity as an operational maintenance criterion. A framework was considered active if its main repository had received at least one commit during the 30 days preceding March 10, 2026. This criterion defines a dated selection snapshot rather than a general assessment of software quality or maintainability.

Seven frameworks met this criterion: Flower, NVIDIA FLARE, FedBioMed, Plato, APPFL, TensorFlow Federated, and XFL. We selected Flower, NVIDIA FLARE, and FedBioMed based on the documented medical-domain applications identified during this screening. This selection establishes the biomedical context of the comparison; the experiments evaluate a specific two-dimensional classification workload rather than the full range of medical-imaging applications.

\subsection{Dataset and Machine Learning Workload}

\begin{figure}[t]
\centering
\includegraphics[width=0.94\linewidth]{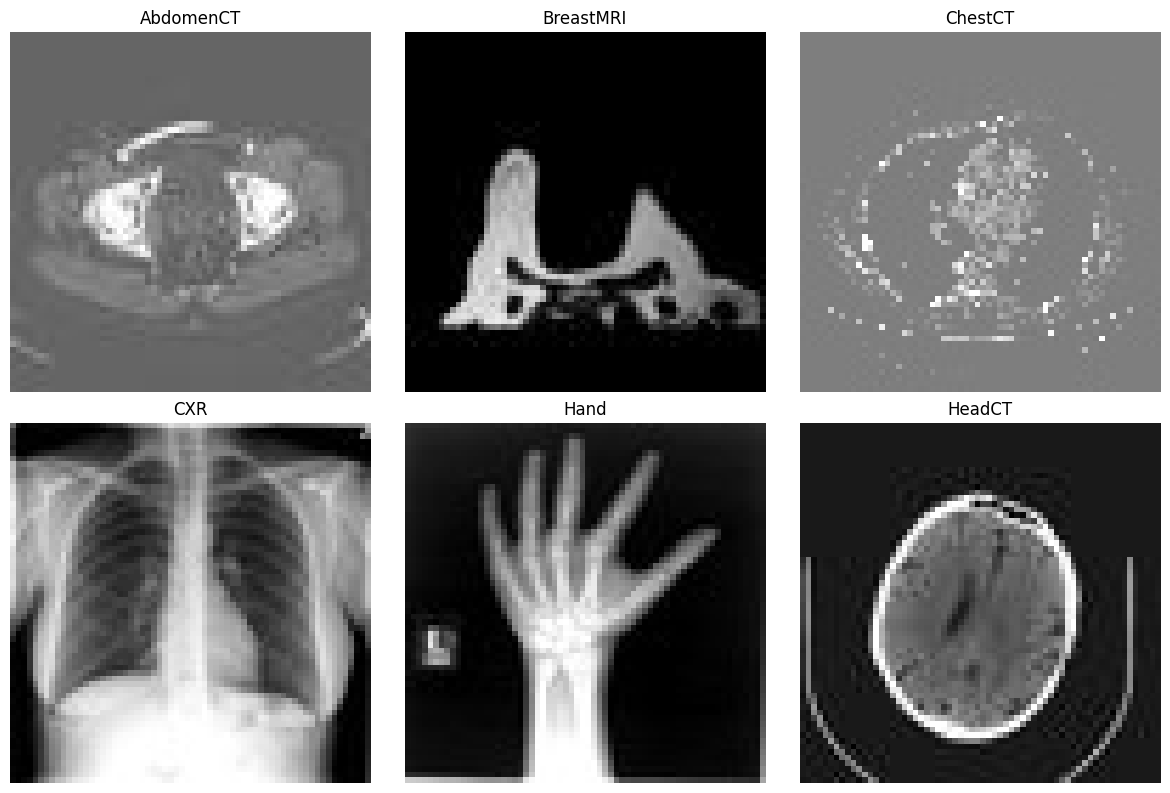}
\caption{Representative examples of the six MedNIST classes: AbdomenCT, BreastMRI, ChestCT, CXR, Hand, and HeadCT.}
\label{fig:mednist}
\end{figure}

The experiments used MedNIST, distributed through the MONAI ecosystem \cite{Cardoso2022}. It contains 58,954 JPEG images from six classes: AbdomenCT, BreastMRI, ChestCT, CXR, Hand, and HeadCT. Figure~\ref{fig:mednist} shows representative images from these six classes. Its relatively small two-dimensional images support repeated resource measurements, while limiting the applicability of the results to larger or volumetric workloads.

Data were partitioned using a pathological non-IID strategy \cite{Li2022}, assigning exactly three classes to each client. The benchmark therefore includes label-distribution heterogeneity under one fixed partitioning strategy, but does not systematically vary the severity of data skew. Inputs were normalized to $[0,1]$. Following the reference MONAI workflow \cite{MONAI2026}, augmentation included rotations of up to $15^{\circ}$, vertical flipping, and zooming between 0.9 and 1.1, each applied independently with probability 0.5.

All implementations used PyTorch and a two-dimensional ResNet-10 \cite{He2016} with one input channel. Training used Adam \cite{Kingma2015}, a learning rate of $10^{-5}$, and cross-entropy loss. These settings define the evaluated workload; the short training schedules were not designed to establish convergence or the best attainable MedNIST accuracy.

\subsection{Federated Training and Evaluation}

Each framework used its native implementation of Federated Averaging (FedAvg), with all clients participating in every round. The nominal local train/test proportion was 80/20. Clients evaluated the global model at the end of each round, and the reported global accuracy was weighted by the number of evaluation samples:

\begin{equation}
 p_t = \sum_{k\in S_t}\left(\frac{n_k}{\sum_{j\in S_t}n_j}\right)p_{kt},
 \label{eq:global_accuracy}
\end{equation}

where $S_t$ is the set of participating clients, $n_k$ is the number of evaluation samples at client $k$, and $p_{kt}$ is its accuracy at round $t$.

Flower and NVIDIA FLARE shared most of the underlying PyTorch code and used the same external component for train/test splitting. FedBioMed required adaptation to its training-plan abstraction and performed this split internally. Consequently, the common split proportion does not establish identical training and evaluation sample membership across all three implementations. Flower also used separate training and federated-evaluation stages, distributing the model in both stages \cite{FlowerEvaluation}. These operational differences are included in the measured execution times and network rates; their individual contributions were not isolated. The results therefore describe these configured workflows rather than a protocol-normalized comparison.

The configured local training budget was two epochs per communication round. This nominal setting should not be interpreted as verification that all framework APIs processed identical minibatches or performed identical numbers of optimizer updates. Results therefore compare the implemented configurations under common high-level settings, with the above differences retained.

\subsection{Experimental Scenarios}

The baseline used $C=3$ clients, $R=3$ rounds, batch size $B=32$, and a nominal local budget of $E=2$ epochs per round. We varied the number of clients, the number of communication rounds, and batch size, as shown in Table~\ref{tab:scenarios}. The $C=10$ configuration used $B=16$ because of GPU-memory limitations. It is retained as a distinct resource-constrained configuration; comparisons involving that point cannot isolate the effect of federation size from the change in batch size.

Each configuration was executed three times using seeds 42, 69, and 420 for each framework. Counting the common baseline once gives nine unique configurations per framework and 81 executions overall.

\begin{table}[t]
\centering
\small
\caption{Experimental configurations. $E$ denotes the nominal local epoch setting.}
\label{tab:scenarios}
\footnotesize\renewcommand{\arraystretch}{1.12}
\begin{tabularx}{\linewidth}{@{}>{\raggedright\arraybackslash}p{0.23\linewidth}>{\raggedright\arraybackslash}p{0.29\linewidth}>{\raggedright\arraybackslash}X@{}}
\hline
Scenario & Varied values & Other settings \\
\hline
Federation size & $C=2,3,5,7$ & $R=3$, $B=32$, $E=2$ \\
Memory-limited & $C=10$ & $R=3$, $B=16$, $E=2$ \\
Rounds & $R=3,5,7$ & $C=3$, $B=32$, $E=2$ \\
Batch size & $B=32,64,128$ & $C=3$, $R=3$, $E=2$ \\
\hline
\end{tabularx}
\end{table}

\subsection{Performance Instrumentation}

Client and server containers \cite{Docker2026} communicated through an internal Docker network. CPU consumption, RAM usage, and network traffic were sampled once per second using cAdvisor \cite{cAdvisor2026}. GPU utilization was collected using NVIDIA DCGM-Exporter \cite{DCGM2026}, and measurements were stored and queried using Prometheus \cite{Prometheus2026}. Server metrics were analyzed separately, while client-side measurements were averaged across participating clients.

GPU utilization was measured for the host rather than for individual containers. It therefore describes activity during the complete execution and cannot attribute GPU use to a particular framework component or distinguish productive computation from effects of concurrent resource use.

Network measurements are reported as rates in KiB/s. A rate depends on the observation interval and execution duration and does not, by itself, establish cumulative bytes transferred. The present analysis does not quantify total communication volume per execution or per round. Similarly, the reported execution times do not separate local training, serialization, communication, orchestration, and aggregation. The measurements characterize overall resource use of the implemented workflows rather than an isolated estimate of framework overhead.

\subsection{Execution Environment}

All experiments ran on the same physical host: an Intel Core i7-13800H processor with 20 threads, 64 GB of DDR5 RAM, an NVIDIA RTX A1000 GPU with 6 GB of VRAM, a 1 TB NVMe SSD, and Ubuntu 22.04 LTS. This controls the available hardware across framework comparisons, while placing all clients and the server in competition for shared resources.

Accordingly, the client-count experiments characterize execution under fixed host resources. They do not establish scalability when clients receive additional dedicated hardware, nor do they reproduce hardware heterogeneity, geographic latency, bandwidth constraints, or intermittent wide-area connections. Observed resource rankings apply to the tested configurations and environment. Version-specific software identifiers are not reported in this manuscript, which limits exact reproducibility and transfer of the findings to other releases.


\subsection{Statistical Analysis}

The analysis used a configuration-blocked design. For each framework and configuration, the three repetitions were first averaged, yielding nine configuration-level observations per framework. Experimental configuration was the blocking factor; the 27 executions per framework were not treated as independent, exchangeable observations for the cross-framework tests.

The Friedman test \cite{Friedman1937} assessed framework differences. When the omnibus result was significant, pairwise comparisons used the Wilcoxon signed-rank test \cite{Wilcoxon1945}, with Holm correction \cite{Holm1979} across the three framework pairs for each metric. Matched-pairs rank-biserial correlation quantified effect size \cite{Tomczak2014}. The significance threshold was $\alpha=0.05$. These tests assess differences across the configurations examined and do not establish universal framework rankings.

Parameter trends are described from the configuration means and variation across repetitions shown in the figures. With only three tested levels for rounds and batch size, these trends are exploratory descriptions; no separate correlation significance tests are used to support scaling or causal claims. The $C=10$, $B=16$ point is explicitly distinguished from the other federation-size configurations.

\section{Results and Discussion}
\label{sec:results}

\subsection{Impact of Federation Size}

\begin{figure*}[t]
\centering
\subfloat[Server transmission rate.]{\includegraphics[width=0.46\textwidth]{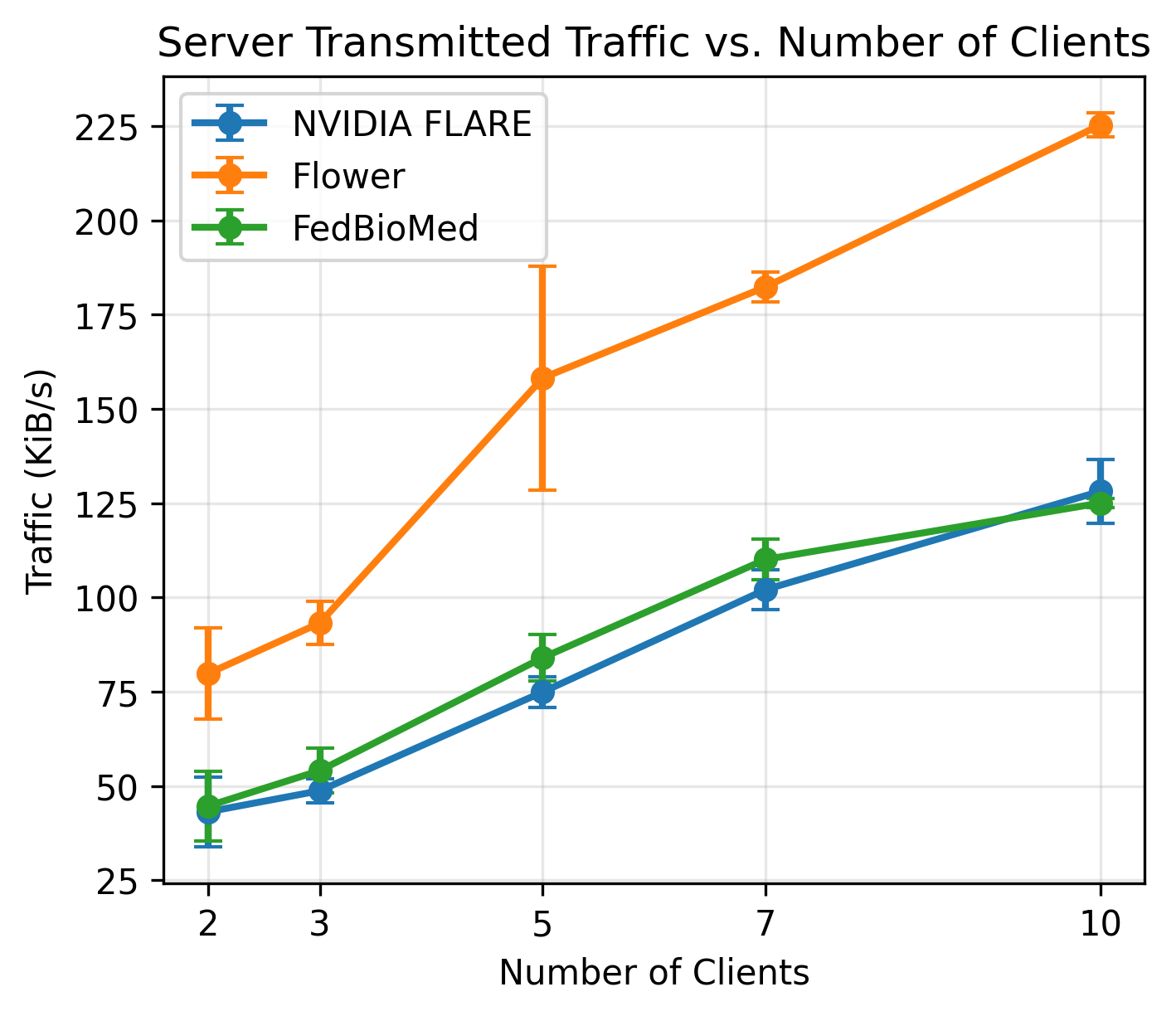}}\hfill
\subfloat[Server reception rate.]{\includegraphics[width=0.46\textwidth]{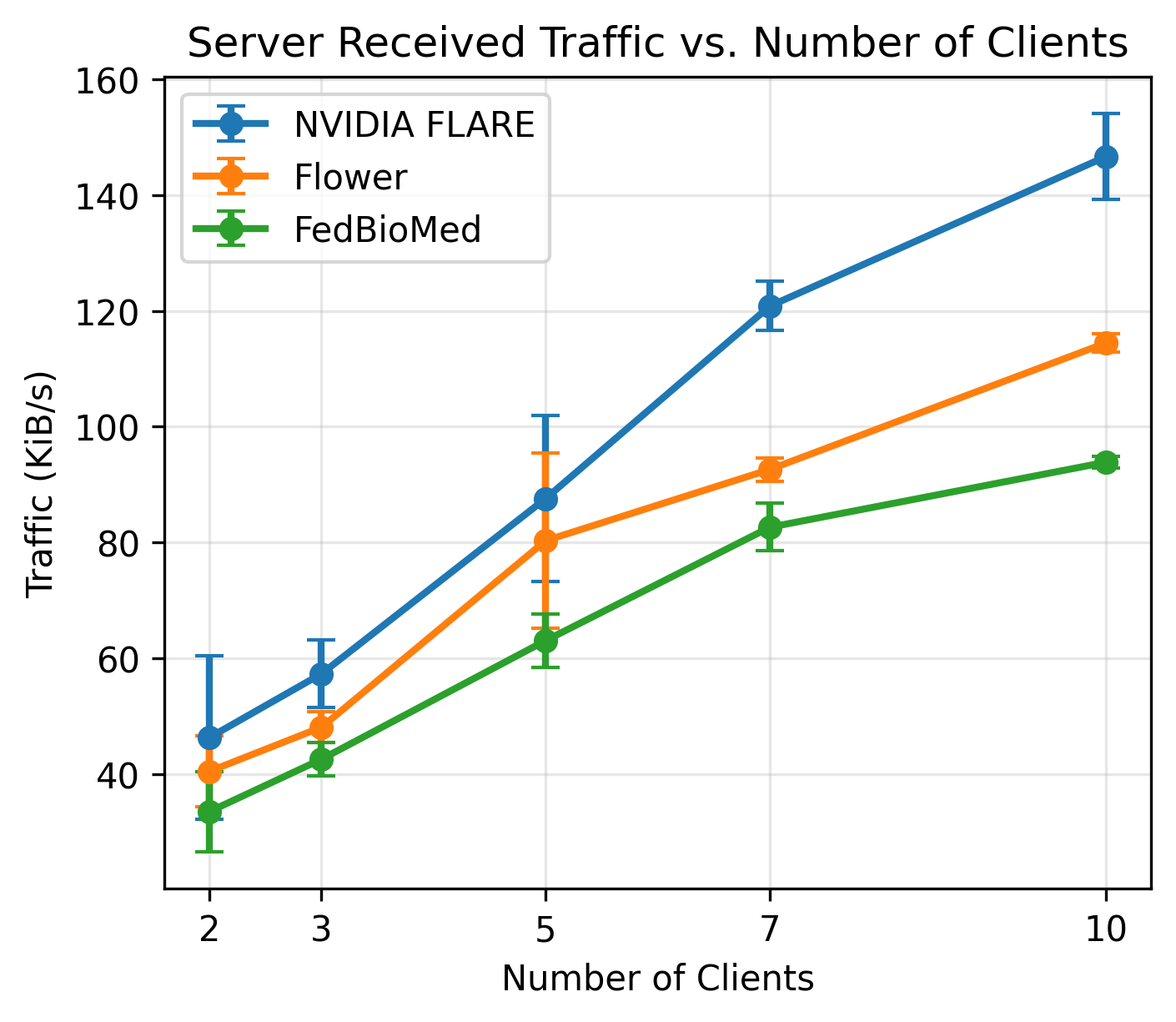}}
\caption{Server-side network rates for the client-count configurations. Error bars show standard deviations across three executions. Rates are expressed in KiB/s. The $C=10$ configuration uses $B=16$; the other configurations use $B=32$. Rates do not measure cumulative transfer volumes.}
\label{fig:network}
\end{figure*}

Across the tested client-count configurations, execution time increased while average CPU usage per client decreased. The total dataset size was fixed, so adding clients reduced the data allocated to each client while increasing the number of concurrent participants on the same host. The measurements are consistent with effects of shared-resource contention and coordination, but do not separate their individual contributions. The $C=10$ result also incorporates the smaller batch size.

Figure~\ref{fig:network} shows the server-side transmission and reception rates. Flower had the highest transmission rate in these configurations, whereas NVIDIA FLARE had the highest reception rate. Flower's separate model distributions for training and evaluation form part of its measured workflow. However, the rate differences do not quantify the volume attributable to that behavior, and a slower execution can exhibit a lower rate without transferring fewer bytes. These results characterize local Docker-network activity rather than communication efficiency over constrained wide-area links.

\subsection{Impact of Communication Rounds}

\begin{figure}[t]
\centering
\includegraphics[width=0.91\linewidth]{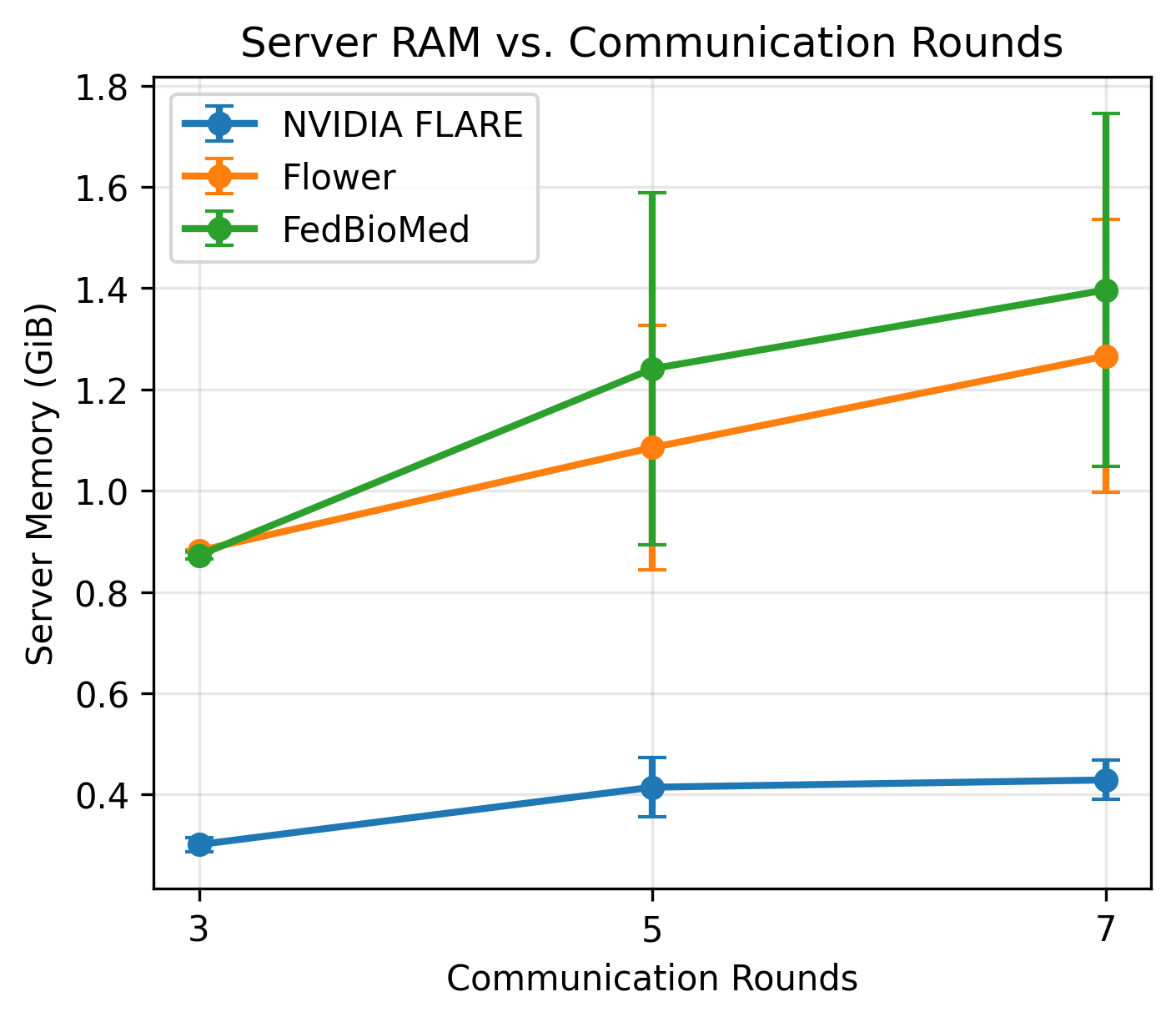}
\caption{Average server RAM for the three evaluated round counts. Error bars show standard deviations across three executions. The figure describes the sampled training budgets and does not establish a long-term memory-growth law.}
\label{fig:rounds_memory}
\end{figure}

Increasing the number of rounds increased total execution time. Figure~\ref{fig:rounds_memory} also shows an increase in average server RAM across the tested round counts, with NVIDIA FLARE remaining below Flower and FedBioMed. This is a descriptive trend across $R=3,5,7$. It does not establish the mechanism responsible for the memory differences or whether memory consumption would continue to grow, stabilize, or change under longer training schedules.

\subsection{Impact of Batch Size}

\begin{figure}[t]
\centering
\subfloat[Execution time in minutes.]{\includegraphics[width=0.85\linewidth]{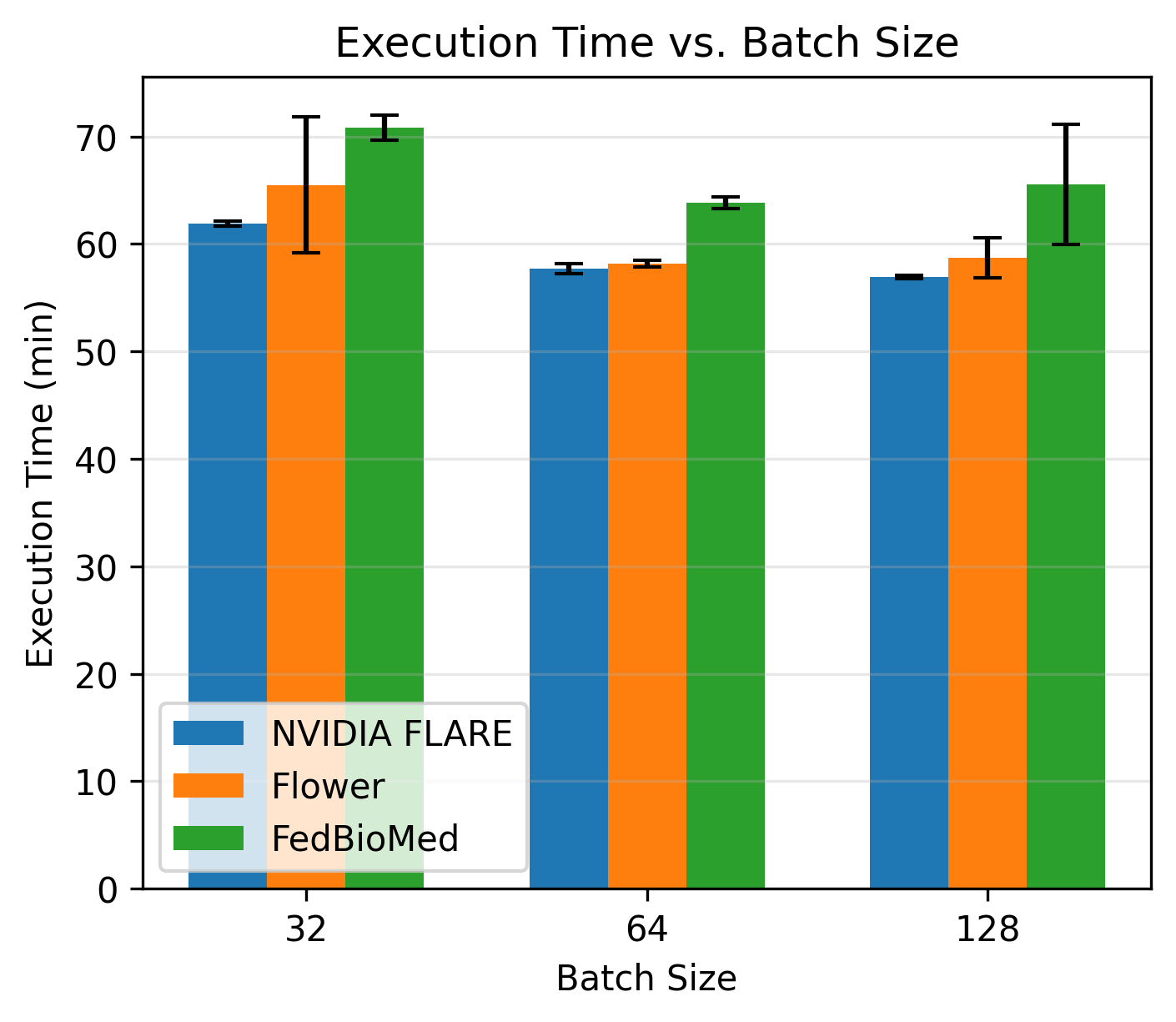}}\\
\subfloat[Host-level GPU utilization.]{\includegraphics[width=0.85\linewidth]{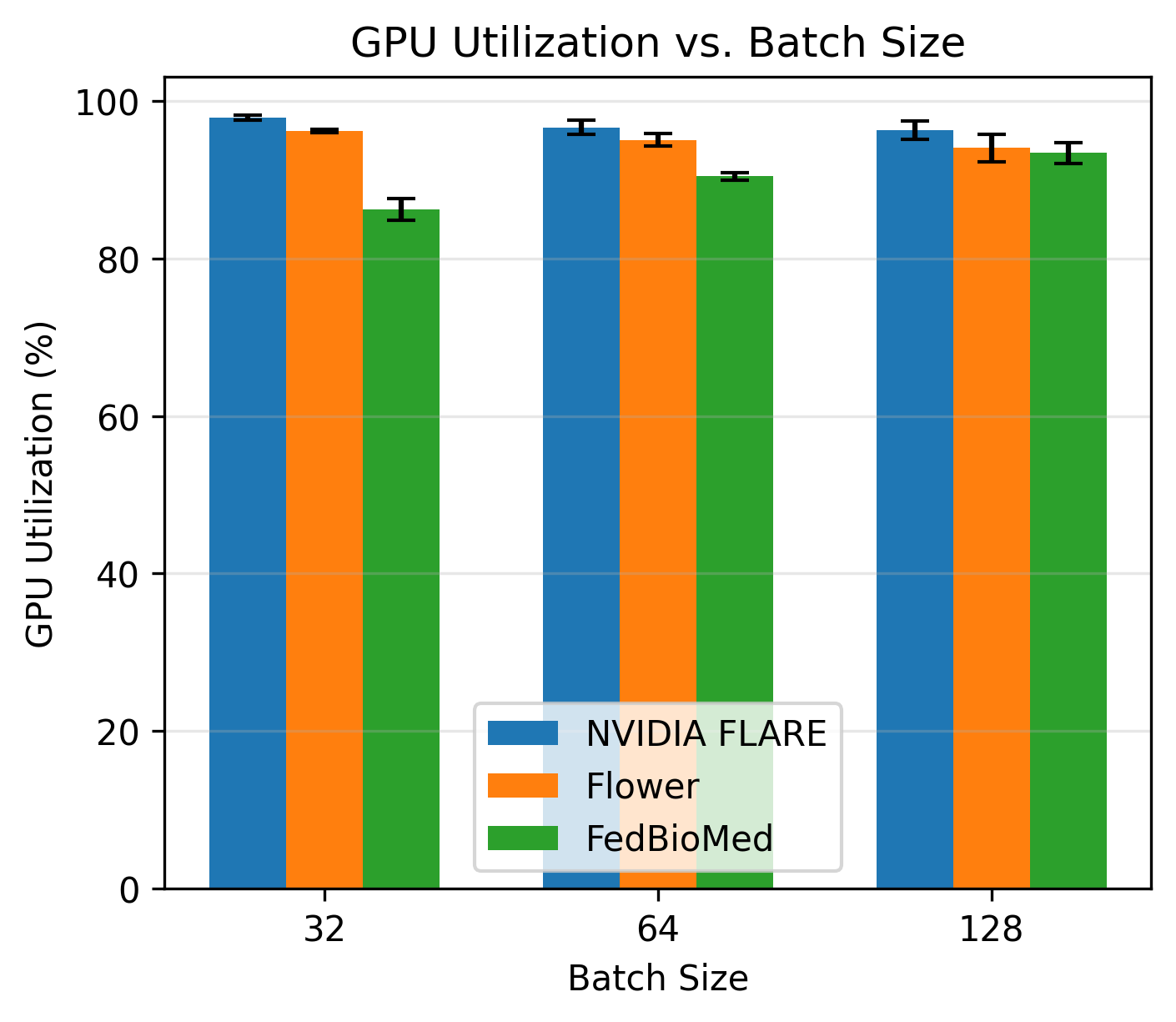}}
\caption{Execution time and host-level GPU utilization at the evaluated batch sizes. Error bars show standard deviations across three executions. GPU utilization describes host activity and is not a measure of computational efficiency.}
\label{fig:batch}
\end{figure}

Figure~\ref{fig:batch} shows that NVIDIA FLARE exhibited the clearest observed reduction in total execution time as batch size increased. Changes were less pronounced for Flower and FedBioMed. These comparisons describe the three tested batch sizes and do not identify a particular internal bottleneck.

Host-level GPU utilization remained near saturation for NVIDIA FLARE. For FedBioMed, it increased from approximately 86\% at $B=32$ to 93\% at $B=128$, without a proportional reduction in total execution time. Higher utilization alone does not establish higher computational efficiency: the measurements do not provide component-level attribution or quantify useful work per unit of GPU activity. The relationship between batch size and total runtime may involve both training and other workflow stages, which were not timed separately.

\subsection{Overall System Performance}

\begin{table*}[t]
\centering
\small
\caption{Descriptive statistics across the 27 executions of each framework (mean $\pm$ standard deviation). Execution times are in minutes. Inferential comparisons use nine configuration-level means.}
\label{tab:system_results}
\begin{tabular}{@{}lrrr@{}}
\toprule
Metric & Flower & NVIDIA FLARE & FedBioMed \\
\midrule
Execution time (min) & $78.55\pm28.81$ & $75.93\pm30.12$ & $89.16\pm32.39$ \\
Client RAM (GiB) & $1.26\pm0.09$ & $1.24\pm0.12$ & $1.66\pm0.10$ \\
Server RAM (GiB) & $1.07\pm0.23$ & $0.35\pm0.06$ & $1.14\pm0.31$ \\
Client CPU (cores) & $1.57\pm0.23$ & $1.55\pm0.23$ & $1.45\pm0.26$ \\
Host GPU utilization (\%) & $96.11\pm1.78$ & $97.53\pm1.64$ & $87.96\pm2.85$ \\
Server network rate (KiB/s) & $184.46\pm74.02$ & $143.17\pm63.55$ & $124.49\pm48.92$ \\
\bottomrule
\end{tabular}
\end{table*}

Table~\ref{tab:system_results} summarizes the descriptive statistics across 27 executions per framework. Cross-framework inference uses the nine configuration-level means described in Section~\ref{sec:experimental_setup}.

The descriptive means show practically relevant resource differences in the tested environment. FedBioMed required approximately 17\% more execution time and 34\% more client RAM than NVIDIA FLARE. NVIDIA FLARE used approximately 67\% less server RAM than Flower. Flower's mean server network rate was approximately 29\% higher than NVIDIA FLARE's and 48\% higher than FedBioMed's. The latter percentages compare rates, not cumulative communication volumes. All these differences describe the complete implemented configurations, including their operational asymmetries.

The Friedman test identified a difference in execution time ($Q=14.889$, $p<0.001$). Holm-corrected Wilcoxon tests showed longer execution times for FedBioMed than for both Flower and NVIDIA FLARE ($p_{\mathrm{adj}}=0.012$, $|r_{\mathrm{rb}}|=1.00$ in both comparisons). Flower and NVIDIA FLARE did not differ significantly ($p_{\mathrm{adj}}=0.250$).

Client RAM also differed ($Q=14.889$, $p<0.001$). FedBioMed used more client RAM than both Flower and NVIDIA FLARE ($p_{\mathrm{adj}}=0.012$, $|r_{\mathrm{rb}}|=1.00$ in both comparisons), whereas the Flower--NVIDIA FLARE comparison was not significant ($p_{\mathrm{adj}}=0.301$). For server RAM ($Q=13.556$, $p=0.001$), NVIDIA FLARE used less memory than both Flower and FedBioMed ($p_{\mathrm{adj}}=0.012$, $|r_{\mathrm{rb}}|=1.00$), while the Flower--FedBioMed comparison was not significant ($p_{\mathrm{adj}}=0.910$). The experiment does not identify which internal mechanisms account for these memory differences.

Client CPU consumption produced a significant omnibus result ($Q=6.222$, $p=0.045$), but no pairwise comparison remained significant after Holm correction. Host-level GPU utilization differed ($Q=14.889$, $p<0.001$): NVIDIA FLARE exceeded Flower ($p_{\mathrm{adj}}=0.020$, $|r_{\mathrm{rb}}|=0.87$) and FedBioMed ($p_{\mathrm{adj}}=0.012$, $|r_{\mathrm{rb}}|=1.00$), and Flower exceeded FedBioMed ($p_{\mathrm{adj}}=0.012$, $|r_{\mathrm{rb}}|=1.00$). These comparisons establish differences in the recorded host-level metric, not superiority in GPU efficiency or utilization by individual containers.

The average server-side network rate also differed ($Q=18.000$, $p<0.001$). Flower's recorded rate exceeded those of NVIDIA FLARE and FedBioMed, and NVIDIA FLARE's exceeded FedBioMed's; all three Holm-corrected pairwise comparisons yielded $p_{\mathrm{adj}}=0.012$ and $|r_{\mathrm{rb}}|=1.00$. These tests concern the rate statistic reported in Table~\ref{tab:system_results} and do not test differences in total bytes per execution or per round.

\subsection{Model Accuracy}

\begin{figure*}[t]
\centering
\includegraphics[width=0.88\textwidth]{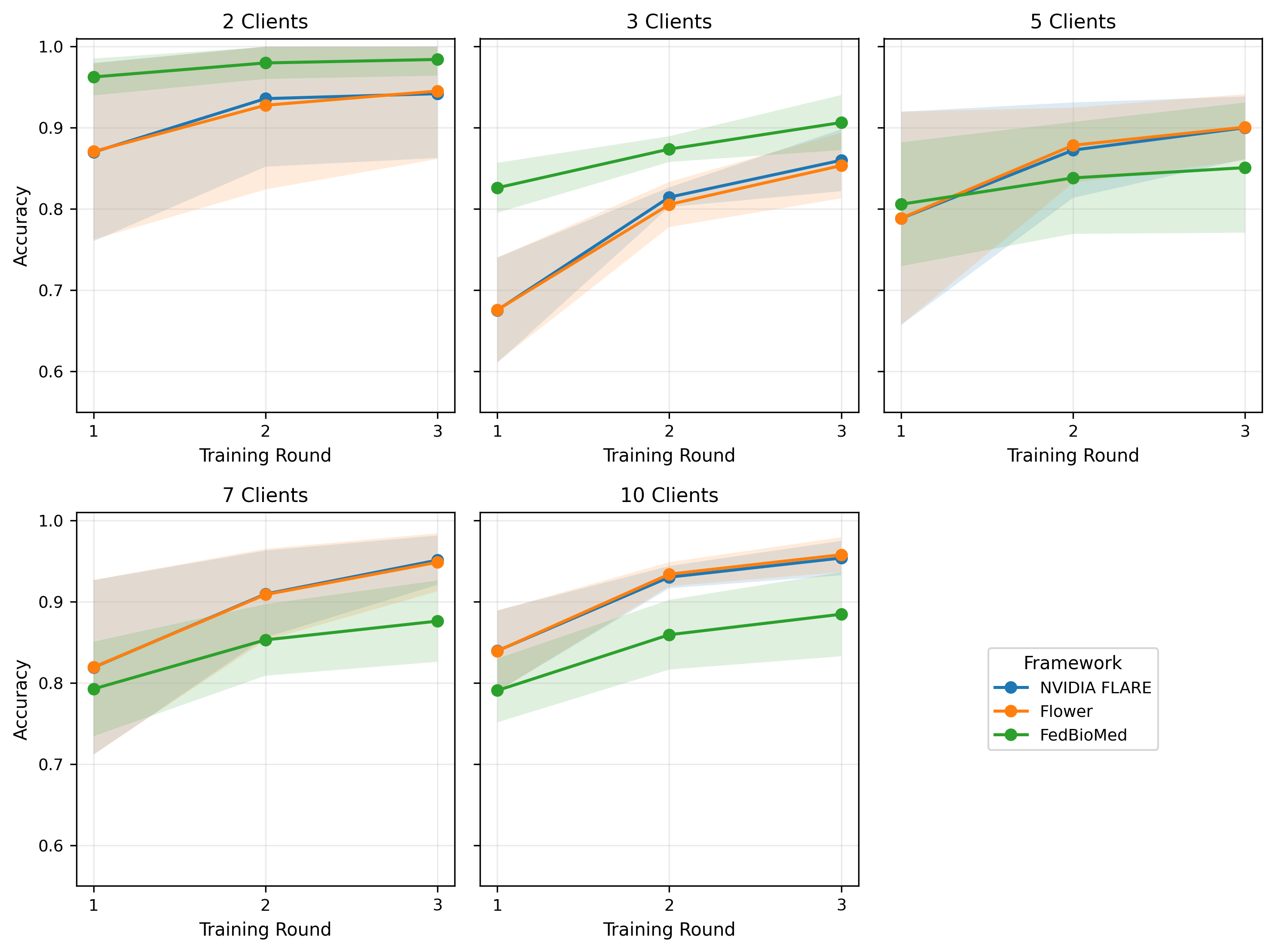}
\caption{Global accuracy over the evaluated training rounds for each federation size. Lines and shaded regions show the mean and one standard deviation across three executions. The $C=10$ configuration uses $B=16$; the other client-count configurations use $B=32$. These short schedules do not establish convergence.}
\label{fig:accuracy}
\end{figure*}

\begin{table}[t]
\centering
\small
\caption{Accuracy at the end of each execution, summarized across all evaluated configurations.}
\label{tab:accuracy}
\begin{tabular}{@{}lcc@{}}
\toprule
Framework & Mean $\pm$ SD & Median \\
\midrule
Flower & $0.8927\pm0.0619$ & 0.8884 \\
NVIDIA FLARE & $0.8946\pm0.0576$ & 0.8753 \\
FedBioMed & $0.9108\pm0.0538$ & 0.9085 \\
\bottomrule
\end{tabular}
\end{table}

Figure~\ref{fig:accuracy} shows accuracy over the short training schedules used in the federation-size experiments. Accuracy serves as an accompanying learning outcome for the resource measurements, without establishing convergence or the best attainable predictive performance.

Across all configurations, final accuracy was $0.8927\pm0.0619$ for Flower, $0.8946\pm0.0576$ for NVIDIA FLARE, and $0.9108\pm0.0538$ for FedBioMed (Table~\ref{tab:accuracy}). The Friedman test did not detect a framework difference ($Q=1.556$, $p=0.459$). This non-significant result does not demonstrate equivalent predictive performance, particularly given the limited repetitions, short schedules, and differences in data splitting and evaluation workflows.

FedBioMed displayed higher accuracy in some smaller-federation configurations. Its internally managed split is one possible contributor, but the experiment does not identify the cause of that pattern. Since training and evaluation sample membership were not established as identical across all frameworks, the observed accuracy values should not be interpreted as a framework-intrinsic learning advantage. The evidence supports a comparison of learning outcomes at the tested training budgets, with longer-term convergence and predictive equivalence remaining unassessed.

\subsection{Practical Implications and Limits}

The measurements identify resource requirements to consider when reproducing this particular workload. NVIDIA FLARE's lower server-memory footprint is relevant when the coordination process faces a memory constraint. FedBioMed's higher client-memory consumption and longer execution times indicate additional resource requirements for its tested configuration. Flower's higher server network rate motivates examining cumulative traffic and the configured evaluation workflow before making bandwidth or communication-cost decisions. Table~\ref{tab:practical_implications} connects these observations to their interpretation limits.

\begin{table*}[t]
\centering
\small
\caption{Measured observations and their practical interpretation within the evaluated configurations. This table does not assess version-specific feature availability.}
\label{tab:practical_implications}
\renewcommand{\arraystretch}{1.10}
\begin{tabularx}{\textwidth}{@{}>{\raggedright\arraybackslash}p{0.17\textwidth}>{\raggedright\arraybackslash}X>{\raggedright\arraybackslash}X@{}}
\hline
Aspect & Observation & Interpretation boundary \\
\hline
Server memory & NVIDIA FLARE used the least server RAM. & Relevant to the tested coordination workload; the responsible internal mechanisms were not isolated. \\
Client resources & FedBioMed used more client RAM and took longer than the other frameworks. & Applies to the implemented workflows, including their training and evaluation differences. \\
Communication & Flower had the highest reported mean server network rate. & Rate alone cannot establish total bytes, bandwidth requirements, or communication efficiency. \\
GPU activity & NVIDIA FLARE had the highest host-level utilization. & Host activity does not identify component usage or establish useful computation per unit of resource. \\
Learning outcome & No significant difference in final accuracy was detected. & Short schedules and workflow differences preclude claims of convergence or predictive equivalence. \\
\hline
\end{tabularx}
\end{table*}

These observations provide a starting point for framework selection under similar conditions. Broader deployment decisions also depend on the actual software versions, required functionality, hardware allocation, network conditions, and workload. The experiment does not rank security mechanisms, governance capabilities, usability, or support for volumetric medical data. Such capabilities require a separate assessment against documentation for the versions to be deployed.

\section{Conclusion}
\label{sec:conclusion}

This work compared Flower, NVIDIA FLARE, and FedBioMed using MedNIST classification with ResNet-10 in a single-host containerized environment. Across 81 executions, the study varied client count, communication rounds, and batch size while monitoring resource consumption and accuracy. The implementations shared nominal learning settings but retained differences in data splitting and federated evaluation.

The configuration-blocked analysis identified substantial differences in the observed resource profiles. FedBioMed required more client RAM and longer execution time than the other frameworks. NVIDIA FLARE exhibited the lowest server-memory consumption and the highest host-level GPU utilization, while Flower had the highest mean server network rate. These findings describe complete configured executions: GPU activity does not establish computational efficiency, and network rates do not establish cumulative communication volume. No statistically significant difference in accuracy was detected under the short training schedules, which does not demonstrate convergence or predictive equivalence.

The parameter experiments provide descriptive evidence of increasing server-memory consumption with round count and framework-dependent responses to batch size. FedBioMed's GPU utilization increased with larger batches without a proportional reduction in total runtime. These observations motivate component-level investigation but do not identify serialization, orchestration, communication, or aggregation as the cause of the differences.

The scope is bounded by shared CPU/GPU resources, local Docker networking, three repetitions per configuration, and a lightweight two-dimensional workload. The $C=10$, $B=16$ configuration changes both client count and batch size. The pathological partitioning introduces label heterogeneity, but the study does not compare controlled degrees of skew or heterogeneous client hardware. Workflow differences, unverified equivalence of effective local update counts, and the absence of release identifiers further limit reproducibility and cross-framework attribution. Consequently, the results support initial resource planning for similar configurations, without establishing deployment-wide rankings for clinical or geographically distributed systems.

Future evaluations should combine verified software and workflow configurations with longer training schedules, more demanding medical workloads, and heterogeneous or distributed participants. Total bytes sent and received per round and separate timings for training, communication, and aggregation would enable more direct analysis of communication volume and execution overhead. These extensions are required to assess whether the observed resource profiles persist beyond the present experimental budget.

\end{document}